\documentclass[viruses,article,submit,LaTeX,moreauthors]{Definitions/mdpi} 
\preto{\abstractkeywords}{\nolinenumbers} 
\firstpage{1} 
\pubvolume{1}
\issuenum{1}
\articlenumber{0}
\pubyear{2022}
\copyrightyear{2022}
\datereceived{} 
\dateaccepted{} 
\datepublished{} 
\hreflink{https://doi.org/} 

\Title{Evaluating the impact of quarantine measures on COVID-19 spread}

\TitleCitation{Title}

\Author{Renquan Zhang $^{1}$\orcidA{}, Yu Wang $^{1}$, Zheng Lv $^{2}$\orcidB{} and Sen Pei $^{3,}$*\orcidC{}}

\AuthorNames{Renquan Zhang, Yu Wang, Zheng Lv and Sen Pei}

\AuthorCitation{Zhang, R.; Wang, Y.; Lv, Z.; Pei, S.}

\address{%
$^{1}$ \quad School of Mathematical Sciences, Dalian University of Technology, Dalian 116024, China; zhangrenquan@dlut.edu.cn (R.Z.), wwy@mail.dlut.edu.cn (Y.W.);\\
$^{2}$ \quad School of Control Science and Engineering, Dalian University of Technology, Dalian 116024, China; lvzheng@dlut.edu.cn (Z.L.);\\
$^{3}$ \quad Department of Environmental Health Sciences, Mailman School of Public Health, Columbia University, New York, NY 10032, USA; sp3449@cumc.columbia.edu (S.P.);}

\corres{Correspondence: sp3449@cumc.columbia.edu (S.P.); }

\abstract{During the early stage of the COVID-19 pandemic, many countries implemented non-pharmaceutical interventions (NPIs) to control the transmission of SARS-CoV-2, the causative pathogen of COVID-19. Among those NPIs, quarantine measures were widely adopted and enforced through stay-at-home and shelter-in-place orders. Understanding the effectiveness of quarantine measures can inform decision-making and control planning during the ongoing COVID-19 pandemic and for future disease outbreaks. In this study, we use mathematical models to evaluate the impact of quarantine measures on COVID-19 spread in four cities that experienced large-scale outbreaks in the spring of 2020: Wuhan, New York, Milan, and London. We develop a susceptible-exposed-infected-removed (SEIR)-type model with a component of quarantine and couple this disease transmission model with a data assimilation method. By calibrating the model to case data, we estimate key epidemiological parameters before lockdown in each city. We further examine the impact of quarantine rates on COVID-19 spread after lockdown using model simulations. Results indicate that quarantine of susceptible and exposed individuals and undetected infections is necessary to contain the outbreak; however, the quarantine rates for these populations can be reduced through faster isolation of confirmed cases. We generate counterfactual simulations to estimate effectiveness of quarantine measures. Without quarantine measures, the cumulative confirmed cases could be 73, 22, 43 and 93 times higher than reported numbers within 40 days after lockdown in Wuhan, New York, Milan, and London. Our findings underscore the essential role of quarantine during the early phase of the pandemic.}

\keyword{COVID-19; Compartmental Model; Quarantine; Data Assimilation} 

\begin{document}


\section*{Introduction}

\label{sec:Introduction}

Emerged in late 2019, a new respiratory pathogen, SARS-CoV-2, rapidly spread across the globe and caused a global pandemic. As of June 14, 2021, more than 176 million confirmed cases have been reported worldwide, of which more than 3.8 million have died \cite{dong2020interactive}. The disease caused by SARS-CoV-2, the coronavirus disease 2019 (COVID-19), is characterized by a substantial proportion of infections with mild/no symptoms \cite{li2020substantial} and a strong age gradient in the risk of death \cite{verity2020estimates, xu2020reconstruction}. During the early stage of the COVID-19 outbreak, the number of confirmed cases generally followed an exponential increase. Studies have shown that the average estimate of the basic reproductive number $R_0$ is between 2.24-3.58 \cite{zhao2020preliminary}. After China implemented strict control measures, the spread of COVID-19 within China was greatly reduced \cite{kraemer2020effect, zhang2019patterns, kucharski2020early, du2020risk}. In other countries, after the initial reporting of infection cases, NPIs such as suspension of classes, cessation of large-scale gatherings, and closure of entertainment and leisure venues have been adopted. These control measures were estimated to effectively slow down the community transmission of SARS-CoV-2 \cite{flaxman2020estimating,brauner2021inferring,hsiang2020effect}.

Before the development of vaccine and its wide administration, NPIs are the primary means to reduce the spread of SARS-CoV-2 \cite{ferguson2020report, ali2020serial, xu2021multiscale}. During vaccination campaign, NPIs also remain key in reducing infections \cite{galanti2020importance}. Among the implemented NPIs, quarantine was used to separate individuals who might have been exposed to COVID-19 from others, which prevents spread of COVID-19 that can occur before a person knows they are infected. During the early days of the COVID-19 pandemic, case isolation and contact tracing were employed to contain the outbreak; however, for an infectious disease whose infectiousness begins before symptoms appear, the effectiveness of isolating cases and tracing contacts is limited \cite{fraser2004factors, peak2017comparing, hellewell2020feasibility}. Indeed, a study found that, for the Lombardy ICU network in Italy, strict self-quarantine measures may be the only possible way to contain the spread of infection \cite{grasselli2020critical}. Despite its effectiveness in disease control, blanket population-level quarantine has enormous economic, social, and educational costs. As a result, understanding the impact of quarantine measures on COVID-19 spread and the intensity of quarantine required to contain an outbreak is critical for planning sustainable control measures by governments and public health authorities. 

In this study, we developed a mathematical model to estimate the effect of quarantine on suppressing COVID-19 spread in four cities: Wuhan in China, New York City in the US, Milan in Italy, and London in the UK. Those cities experienced early outbreaks of COVID-19 and all enforced strict interventions to control the transmission of SARS-CoV-2. We incorporated a component of quarantine into a classical susceptible-exposed-infected-removed (SEIR) model, and calibrated the model to confirmed cases in each city during the early phase of the pandemic using a data assimilation method. We estimated key epidemiological parameters before lockdown in each city, and evaluated the impact of the quarantine rates of susceptible, exposed and undetected infected populations on disease transmission. Particularly, we estimated the required minimal quarantine rates of those populations to reduce the effective reproductive number below 1 at the beginning of lockdown. We further simulated counterfactual outbreaks within 40 days following lockdown, assuming no quarantine were implemented in those cities, and estimated the averted cases attributed to quarantine measures. Overall, quarantine measures have effectively prevented 3,674,995, 3,512,956, 649,880 and 2,285,718 reported cases during this 40-day period in Wuhan, New York, Milan, and London, respectively.
\section{Methods}
\label{sec:Method}

\subsection{Model}
\label{subsec:Model}

\begin{figure}[H]
\centering
\includegraphics[width=10.5 cm]{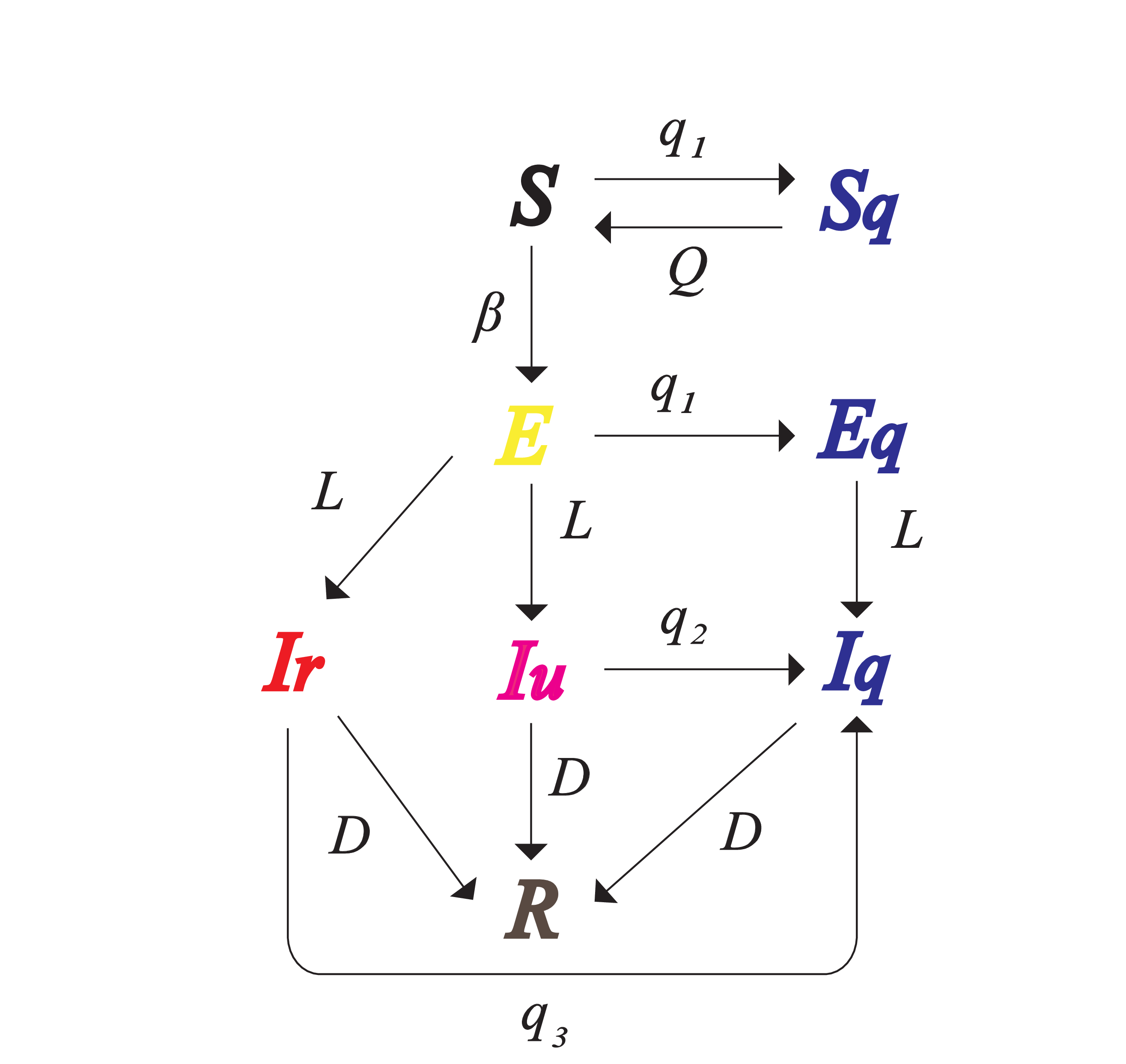}
\caption{Dynamics of the transmission model. The compartments $S$, $E$, $I_r$, $I_u$ and $R$ represent susceptible, exposed, reported infected, unreported infected and removed populations. $S_q$, $E_q$ and $I_q$ are susceptible, exposed and infected individuals under quarantine. $q_1$ is the quarantine rate of susceptible and exposed persons, $q_2$ is the quarantine rate of unreported infections, and $q_3$ is the isolation rate of confirmed cases. $\beta$ is the transmission rate of SARS-CoV-2, $L$ is the mean duration of latency period, $D$ is the mean duration of infectious period, and $Q$ is the average length of quarantine.}
\label{figure1:model}
\end{figure}   

We used a modified SEIR model to depict the transmission of SARS-CoV-2 in a location with quarantine measures. The model dynamics is shown in Figure \ref{figure1:model}. Specifically, $S$, $E$, $S_q$, $E_q$, $I_r$, $I_u$, $I_q$ and $R$ represent susceptible, exposed, quarantined susceptible, quarantined exposed, reported infected, unreported infected, isolated infected and removed (recovered or dead) populations. A susceptible individual can be infected by a reported infection with a transmission rate $\beta$ or an unreported infection with a transmission rate $\mu\beta$ where $\mu\in[0,1]$. Note we assume undocumented infections are less contagious than confirmed cases, as indicated in previous studies \cite{sayampanathan2021infectivity,bi2020household,gao2020study,byambasuren2020estimating}. Exposed individuals become contagious after a mean latency period of $L$ days. A fraction of infected population, $\alpha$, is ascertained as confirmed cases. Infected individuals recover or die after a mean infectious period of $D$ days. Susceptible and exposed individuals are quarantined at a quarantine rate $q_1$, while unreported infections are quarantined at a rate $q_2$. Here quarantine rate is defined as the probability of an individual to be quarantined on a given day. A single quarantine rate $q_1$ is defined for both susceptible and exposed individuals as they are typically indistinguishable without indication of illness. We used two different quarantine rates to reflect the differential perception of infection risk for undocumented cases who may have mild symptoms. Confirmed cases are isolated following an isolation rate $q_3$ after they become infectious. Quarantined susceptible people are released from quarantine after $Q$ days. We assume quarantined individuals (susceptible, exposed or infected) do not participate in disease transmission. The transmission dynamics is described by the following equations:
\begin{align}
&\frac{dS}{dt}=-q_1S+(1-q_1)\left(\frac{\beta I_rS}{N}+\frac{\mu \beta I_uS}{N}\right)+\frac{S_q}{Q} \label{eq1}\\
&\frac{dE}{dt}=(1-q_1)\left(\frac{\beta I_rS}{N}+\frac{\mu \beta I_uS}{N}\right)-q_1E-(1-q_1)\frac{E}{L} \label{eq2}\\
&\frac{dE_q}{dt}=q_1E-\frac{E_q}{L} \label{eq3}\\
&\frac{dS_q}{dt}=q_1S-\frac{S_q}{Q} \label{eq4}\\                   
&\frac{dI_r}{dt}=\alpha (1-q_1)\frac{E}{L}-\frac{I_r}{D}-q_3I_r \label{eq5}\\
&\frac{dI_u}{dt}=(1-\alpha)(1-q_1)\frac{E}{L}-\frac{I_u}{D}-q_2I_u \label{eq6}\\
&\frac{dI_q}{dt}=\frac{E_q}{L}-\frac{I_q}{D}+q_2I_u+q_3I_r \label{eq7}\\
&\frac{dR}{dt}=\frac{I_r}{D}+\frac{I_u}{D}+\frac{I_q}{D} \label{eq8}
\end{align}
Using model equations, we compute the effective reproductive number, $R_e$, i.e. the average number of new infections caused by a single infected individual in a population with partial immunity, as
\begin{equation}
R_e=\left(\frac{(1-q_1)\alpha \beta}{q_3+\frac{1}{D}}+\frac{(1-q_1)(1-\alpha) \mu \beta}{q_2+\frac{1}{D}}\right)\frac{S}{N}.
\label{equation10:effective}
\end{equation}
In model simulations, we deterministically integrate equations using the 4th-order Runge-Kutta method.

\subsection{Model calibration}
\label{subsec:Inference}

We calibrated the transmission model to daily confirmed cases in each city using a data assimilation method - the ensemble adjustment Kalman filter (EAKF) \cite{anderson2001ensemble}. The EAKF is a recursive filtering technique that assimilates observations into a dynamic model to generate a posterior estimate of model state. It has been widely used in numerical weather prediction \cite{anderson2001ensemble,navon2009data} as well as inference and forecasting of infectious diseases such as influenza \cite{pei2018forecasting}, COVID-19 \cite{pei2021burden}, and other respiratory viruses \cite{pei2021optimizing}.

The EAKF assumes a Gaussian distribution of both the prior and likelihood and adjusts the prior distribution to a posterior using Bayes’ rule. To represent the state-space distribution, the EAKF maintains an ensemble of system state vectors acting as samples from the distribution. In particular, the EAKF assumes that both the prior distribution and likelihood are Gaussian, and thus can be fully characterized by their first two moments (mean and variance). The update scheme for ensemble members is computed using Bayes’ rule (posterior $\propto$ prior $\times$ likelihood) via the convolution of the two Gaussian distributions. In the EAKF, variables and parameters are updated deterministically such that the higher moments of the prior distribution are preserved in the posterior. Details on the implementation of the EAKF can be found in published studies \cite{anderson2001ensemble,yang2014comparison}.

In the analysis, we first focus on the period before lockdown and stay-at-home order were announced in each city. For Wuhan, New York, Milan and London, we used daily case data reported from January 16 to January 23, March 1 to March 19, February 25 to March 8, and March 6 to March 23, respectively. We collected the daily reported case data in the four cities (see details in Appendix~\ref{sec:data}). To mitigate the impact of possible irregular reporting during the early stage of the pandemic, we used a five-day moving average to smooth the epidemic curves. Before lockdown, confirmed cases were isolated but population-level quarantine was not yet affected. We therefore set the quarantine rates $q_1$ and $q_2$ as zero. To reduce the number of unknown parameters, we fixed several parameters estimated in previous studies. Specifically, the quarantine rate of confirmed cases $q_3$ is estimated as the reciprocal of the delay from the onset of contagiousness to confirmation. For instance, the confirmation delay in Wuhan was estimated to be 6-10 days \cite{li2020substantial,li2020early}; we therefore set the range of $q_3$ to be [0.1, 0.16]. Confirmation delays in New York, Milan and London were also reported in previous studies \cite{pei2020differential,hellewell2020feasibility,li2020early}. We further fixed relative transmission rate $\mu$, the mean latency period $L$ and mean infectious period $D$ as reported in other studies \cite{li2020substantial}. In model simulations and inference, these parameters were uniformly drawn from the prior range and were fixed throughout the analysis. We used the model-inference framework to estimate two parameters: $\alpha$, $\beta$. The fixed ranges of $q_3$, $\mu$, $L$ and $D$ and the prior ranges of $\alpha$, $\beta$ were shown in Table \ref{table1:parameter range} \cite{li2020substantial,pei2020differential,pei2020initial}. We used 300 ensemble members in the EAKF, and drew initial parameters uniformly from the prior ranges. In model simulations, we fixed the average quarantine length $Q$ as 14 days.

\begin{table}[H] 
\caption{\label{table1:parameter range}Prior parameter range for Wuhan, New York, Milan, and London before lockdown.}
\newcolumntype{C}{>{\centering\arraybackslash}X}
\begin{tabularx}{\textwidth}{CCCCCC}
\toprule
  & \textbf{Wuhan}	& \textbf{New York}	& \textbf{Milan} & \textbf{London} & \\
\midrule
$q_3$ & (0.1,0.16) & (0.07,0.11) & (0.1,0.16) & (0.1,0.16) & fixed \\
$\beta$ & (0.6,1.5) & (0.6,1.5) & (0.6,1.5) & (0.6,1.5) & estimated \\
$\mu$ & (0.45,0.65) & (0.45,0.65) & (0.45,0.65) & (0.45,0.65) & fixed\\
$\alpha$ & (0.02,0.2) & (0.02,0.2) & (0.02,0.2) & (0.02,0.2) & estimated\\
L&(3,5)&(3,5)&(3,5)&(3,5)&fixed\\
D&(3,5)&(3,5)&(3,5)&(3,5)&fixed\\
\bottomrule
\end{tabularx}
\end{table}

\section{Results}
\label{sec:Results}

\subsection{Epidemiological characteristics before lockdown}
\label{subsec:Epidemiological}

\begin{figure}[H]
\centering
\includegraphics[width=10.5 cm]{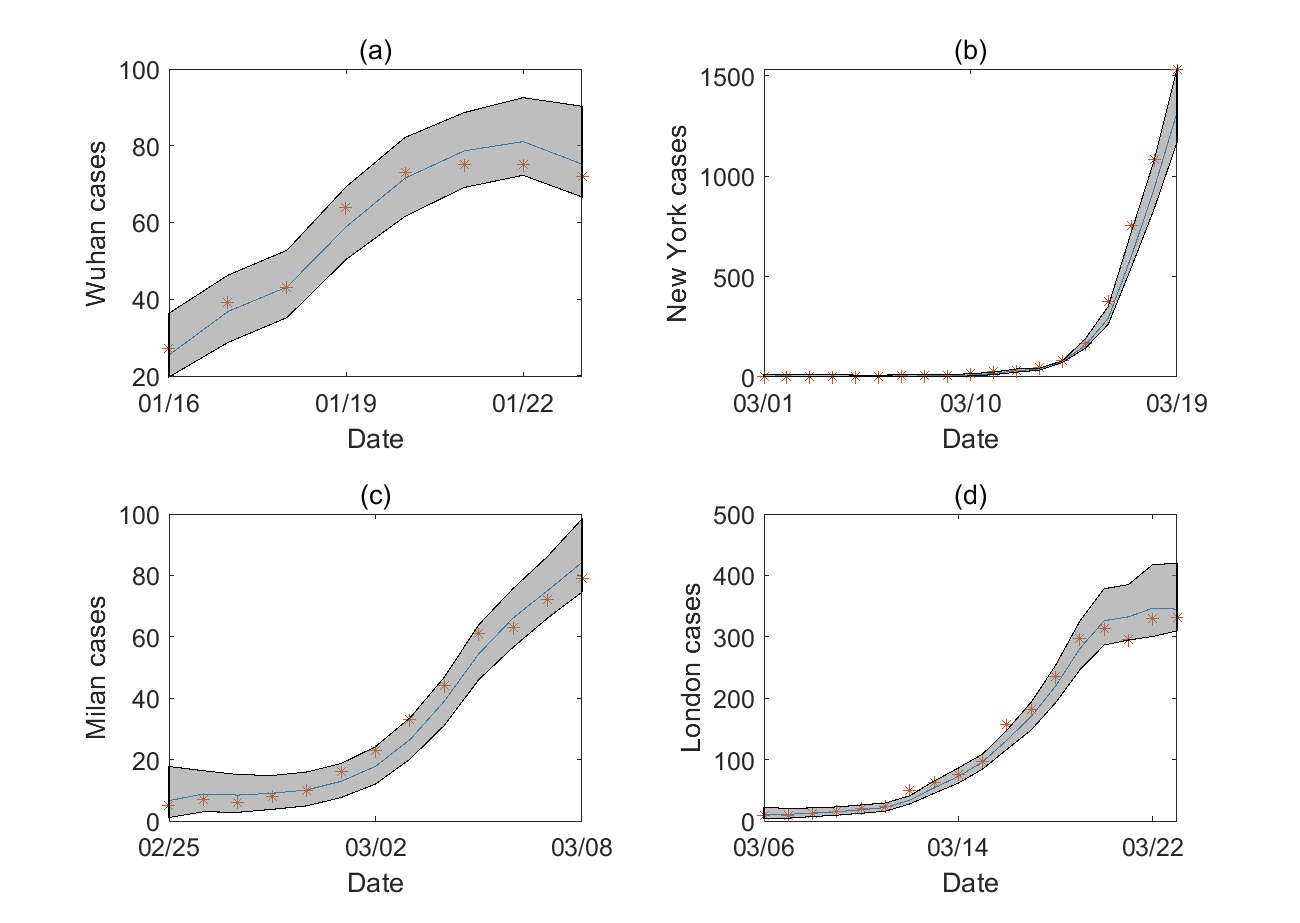}
\caption{Model fitting for (a) Wuhan from January 16 to January 23, (b) New York from March 1 to March 19, (c) Milan from February 25 to March 8, and (d) London from March 6 to March 23. The orange star symbol represents reported case number, the blue curve is the mean posterior fitting using the EAKF, and the gray region shows the 95\% CI.}
\label{figure2:simulation}
\end{figure}

Using the model-data assimilation framework, we estimated the posterior epidemiological parameters in the four cities before lockdown (see Table \ref{table2:parameter value}). The posterior fitting agrees well with observed daily cases, as shown in Figure \ref{figure2:simulation}. Before lockdown, the number of new cases increased rapidly during the study period, and most data points fall within the 95\% CI. The estimated effective reproductive number $R_e$ is generally in line with previous estimates \cite{zhao2020preliminary}. New York has the highest estimated $R_e=3.19$, followed by London ($R_e=2.97$), Milan ($R_e=2.73$) and Wuhan ($R_e=2.26$). Only $8.9\%$ infections were estimated to be confirmed in New York, agreeing with previous modeling results \cite{pei2020initial} and surveys of healthcare-seeking behavior \cite{reese2021estimated} and seroprevalence \cite{stadlbauer2021repeated}. Other cities have slightly higher ascertainment rates, but the majority of infected individuals went undetected due to mild/no symptoms and limited testing capacity. 

\begin{table}[H] 
\caption{\label{table2:parameter value}Posterior parameter estimates of Wuhan, New York, Milan, and London before lockdown.}
\newcolumntype{C}{>{\centering\arraybackslash}X}
\begin{tabularx}{\textwidth}{CCCCC}
\toprule
  & \textbf{Wuhan}	& \textbf{New York}	& \textbf{Milan} & \textbf{London}\\
\midrule
$\beta$&0.99&1.43&1.23&1.31 \\
$\alpha$&0.189&0.089&0.080&0.148\\
$R_e$&2.26&3.19&2.73 &2.97\\
95\% CI & (1.63,3.02) & (2.26,4.28) & (1.85,3.76) & (2.20,3.94) \\
\bottomrule
\end{tabularx}
\end{table}

\subsection{Impact of quarantine rates on COVID-19 spread}
\label{subsec:Change}

\begin{figure}[H]
\includegraphics[width=10.5 cm]{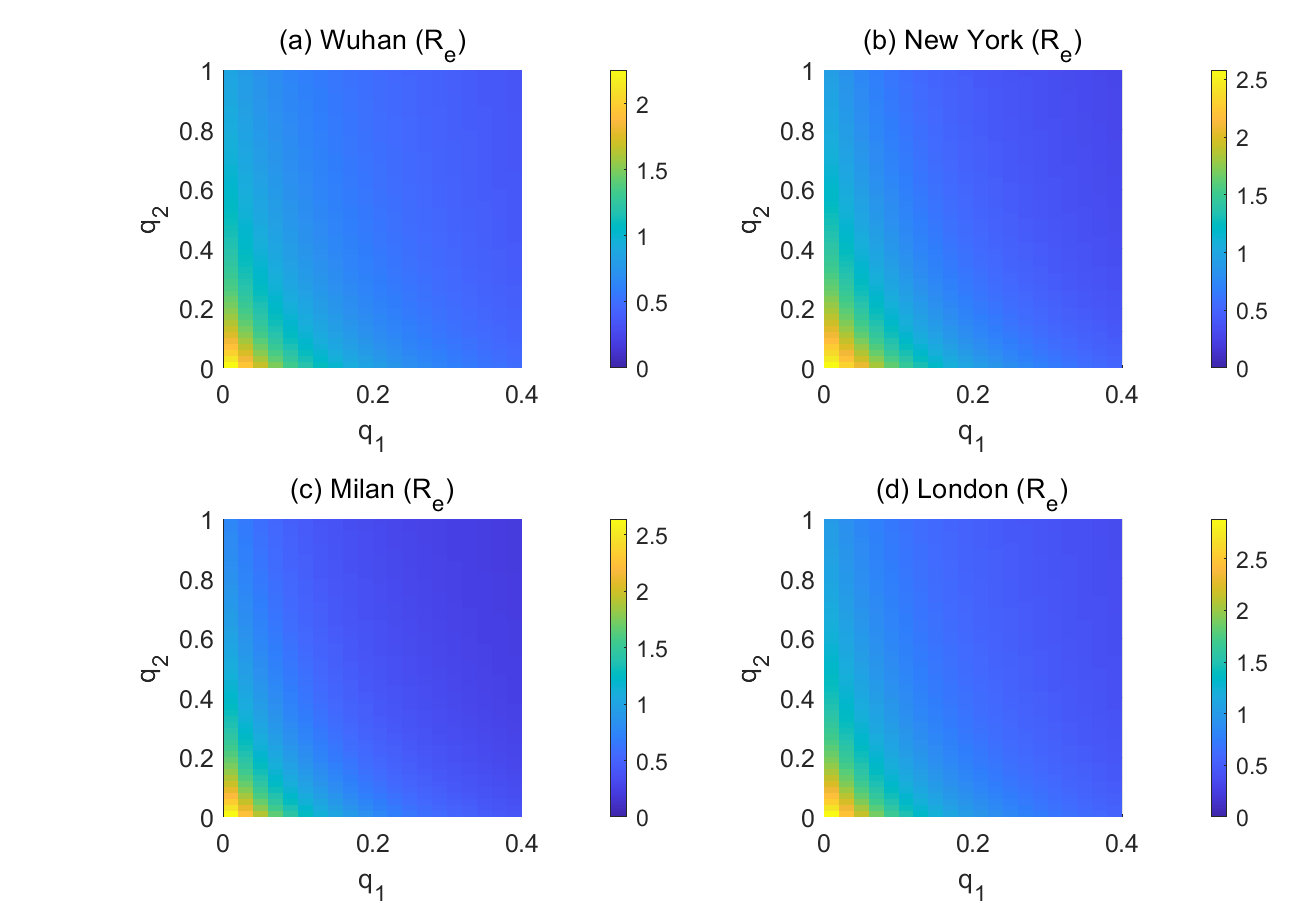}
\centering
\includegraphics[width=9.5 cm]{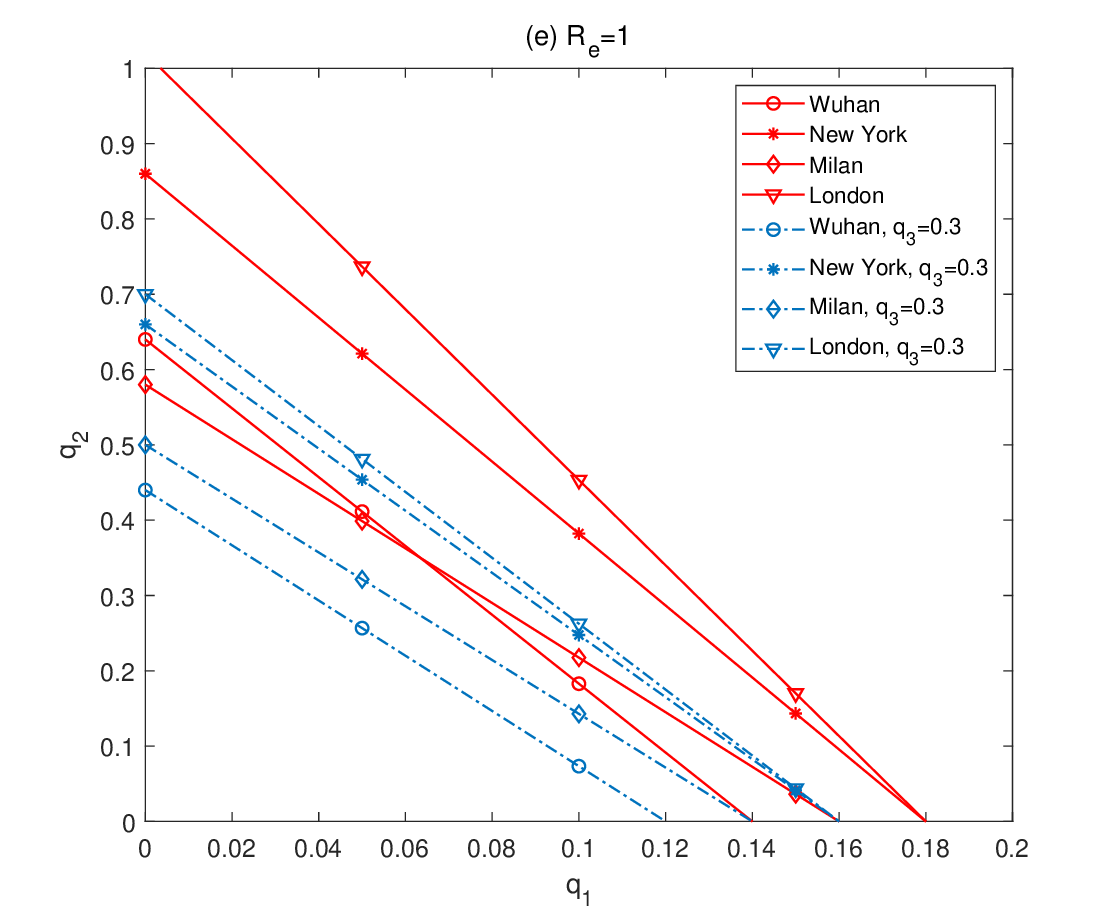}
\caption{(a), (b), (c), (d) show $R_e$ values of Wuhan, New York, Milan, and London at the beginning of model simulation with different combinations of the quarantine rates $q_1$ and $q_2$. Other parameters are set as in Table \ref{table1:parameter range}, \ref{table2:parameter value}. Inset (e) shows the combination of $q_1$ and $q_2$ that lead to $R_e=1$ in the four cities. The solid lines represent results obtained for $q_3$ set as in Table \ref{table1:parameter range}, while the dash lines are the results for $q_3=0.3$.}
\label{figure3:re}
\end{figure}

We use model simulations to examine the minimal quarantine rates required to reduce $R_e$ below one. Specifically, we initialized the transmission model using model states and parameters inferred on the last day before lockdown (as shown in Table \ref {table2:parameter value}), varied the quarantine rates $q_1$ (the quarantine rate of susceptible and exposed individuals) and $q_2$ (the quarantine rate of unreported infections), and ran model simulations until the outbreak stops. We examined the effects of different combinations of quarantine rates on the effective reproductive number, the duration of outbreak, the attack rate, and peak timing.

Figure \ref{figure3:re} shows the impact of quarantine rates $q_1$ and $q_2$ on the effective reproductive number $R_e$ on the first day of model simulation. In order to reduce $R_e$ below 1, undocumented infections need to be quarantined with a much faster quarantine rate than susceptible population. This pattern consistently holds across all four cities. Figure \ref{figure3:re}(e) shows the combinations of $q_1$ and $q_2$ that lead to $R_e=1$ in the four cities. The solid lines represent results obtained for $q_3$ set as in Table \ref{table1:parameter range}, i.e. the same isolation rate of confirmed cases as before lockdown. If control measures on confirmed cases remain the same after lockdown, quarantining only unreported infections is not sufficient to contain the outbreak in London - even with $q_2=1$, the effective reproductive number $R_e$ is above one. For Wuhan, New York and Milan, it is possible to reduce $R_e$ below one through the quarantine of only undetected infections, but the majority of undocumented cases need to be quarantined quickly ($q_2$ is close to 1). In reality, this is very challenging because rapid testing is not widely available and the turnaround time of PCR testing is too long to support timely quarantine. As a result, it is necessary to quarantine susceptible population in order to control the outbreak.

After lockdown, if the isolation rate of confirmed cases $q_3$ increases, the quarantine rates $q_1$ and $q_2$ required to contain the outbreak can be relaxed. In Figure \ref{figure3:re}(e), we use dash lines to show the combinations of $q_1$ and $q_2$ for $R_e=1$ when the isolation rate $q_3$ is 0.3 (i.e., on average, confirmed cases are isolated 3.3 days after they become contagious). With faster isolation of confirmed cases, the outbreak can be contained by quarantining only undocumented infections in all four cities. However, New York and London still need higher quarantine rates than Milan and Wuhan. The required quarantine rate of susceptible population $q_1$ decreases with increased quarantine rate of undocumented infection $q_2$. In order to minimize the population size under quarantine and reduce the disturbance on society, the best control strategy should be to isolate confirmed cases and individuals who are exposed to infections (possible undocumented infections) as soon as possible so that the required quarantine rate of susceptible population could be lower.

\begin{figure}[H]
\centering
\includegraphics[width=10.5 cm]{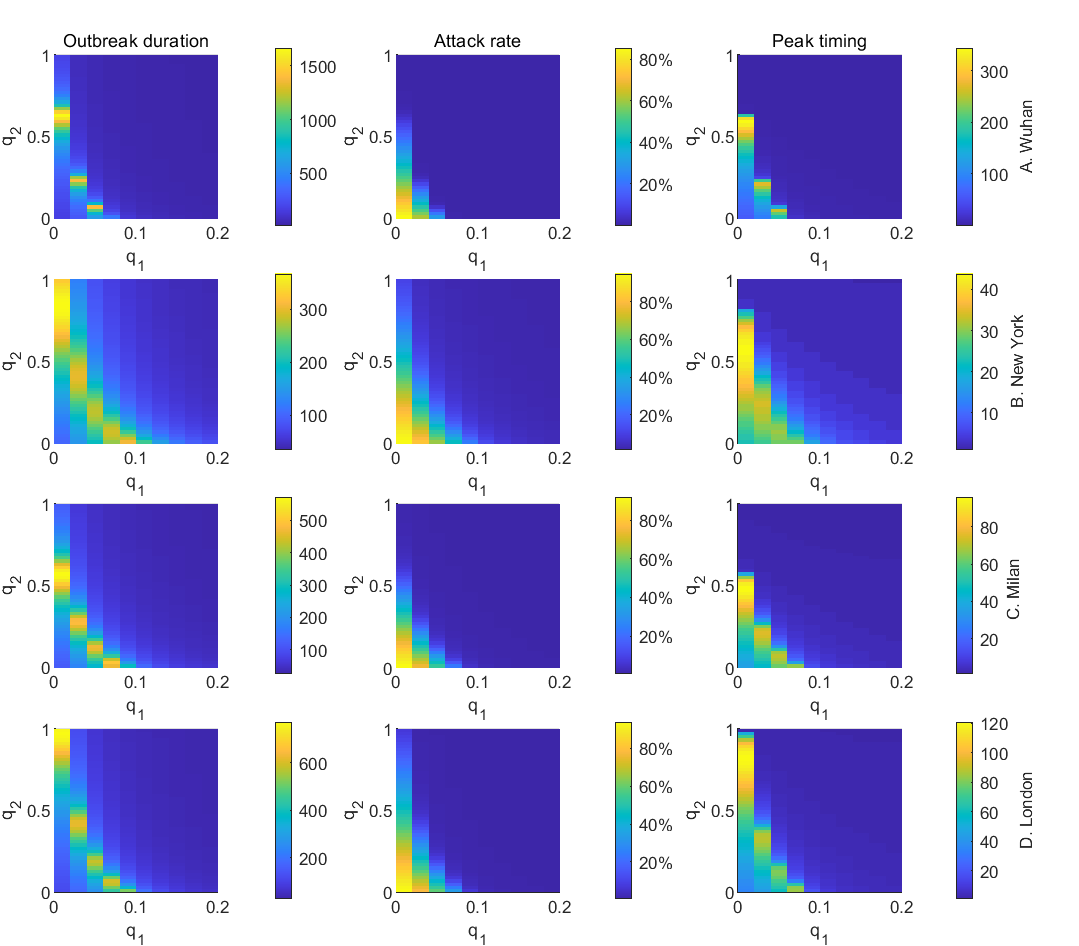}
\caption {Impact of quarantine rates $q_1$ and $q_2$ on the outbreak duration, attack rate, and peak timing. The first column shows the duration of the outbreak, i.e., the number of days it takes for daily cases to drop below 5 after the lockdown measures are enforced; the second column shows the attack rate by the time when daily cases fall below 5; the third column shows peak timing of daily cases after lockdown, defined as the number of days between lockdown and the day with the highest reported daily case. Each row corresponds to one city.}
\label{figure4}
\end{figure}
We further explore the impact of quarantine rates $q_1$ and $q_2$ on several characteristics of the outbreak, including the outbreak duration, attack rate, and peak timing. Here the outbreak duration is defined as the number of days it takes for daily cases to drop below 5 after the lockdown measures are enforced; the attack rate is the percentage of population infected with SARS-CoV-2 by the time daily cases fall below 5; and peak timing is the number of days between lockdown and the day with the highest reported daily case. We ran model simulations using different combinations of $q_1$ and $q_2$ starting from lockdown with other parameters set as in Table \ref{table1:parameter range}, \ref{table2:parameter value}, until the daily case number falls to 5. Simulation results are shown in Figure \ref{figure4}.

The outbreak duration is maximized for the combinations of $q_1$ and $q_2$ that lead to $R_e=1$. For $R_e>1$, the outbreak depletes susceptible population and stops due to herd immunity; for $R_e<1$, the outbreak dies out as the low secondary infection rate cannot support self-sustaining transmission, leaving the majority of population susceptible. At the critical state $R_e=1$, the outbreak would linger for a long period until herd immunity stops disease spread. Peak timing also follows the same pattern, as shown in the right column of Figure \ref{figure4}. The outbreak duration and peak timing is shorter in cities with higher $R_e$ before lockdown. Attack rate increases with lower rates $q_1$ and $q_2$. Without control, over 80\% population in the four cities would be infected.

\subsection{Estimating the averted cases due to quarantine measures}
\label{subsec:Comparison}

\begin{figure}[H]
\centering
\includegraphics[width=10.5 cm]{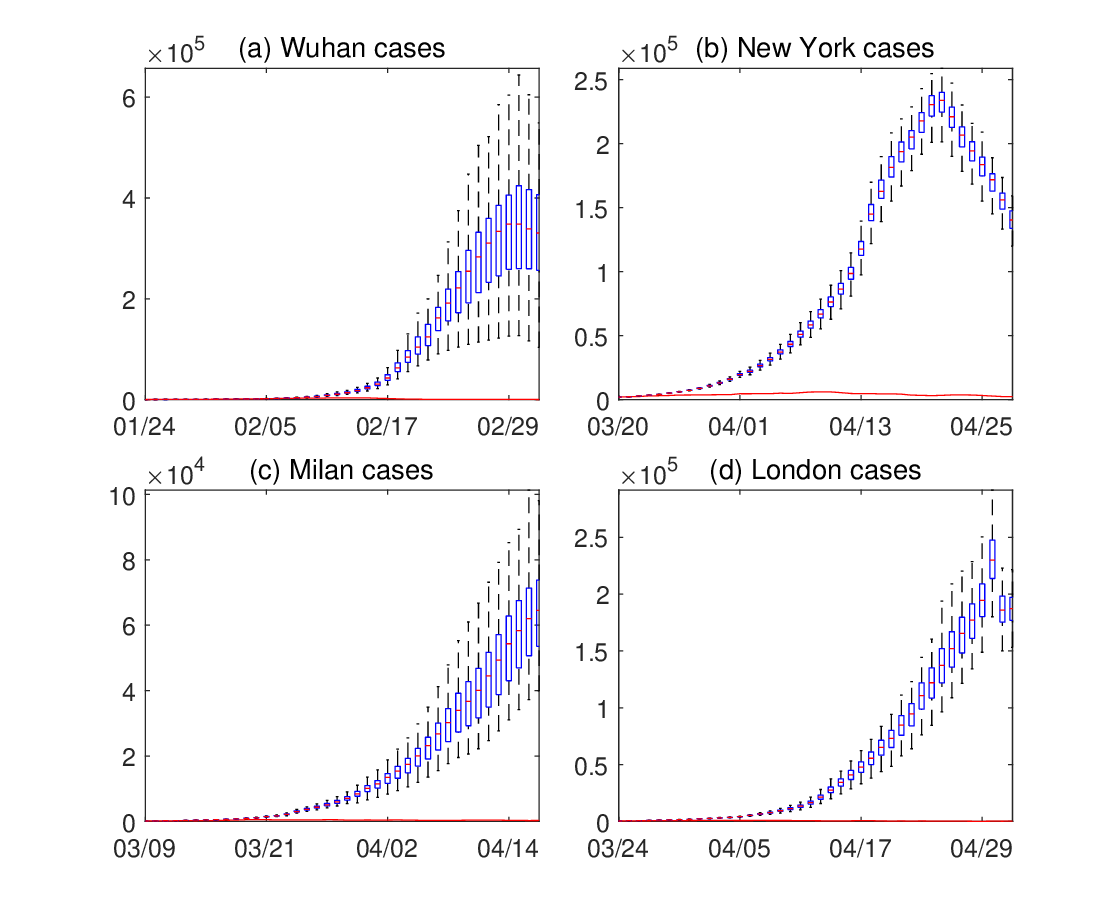}
\caption {Counterfactual simulations of outbreaks in Wuhan, New York, Milan and London. Simulations were performed using posterior model parameters estimated each day in the four cities within 40 days after lockdown, with the quarantine rate $q_1=q_2=0$. The red lines are the observed daily case numbers. Blue boxes show the median and interquartile of counterfactual simulations, and whiskers show 95\% CI. Counterfactual simulations were performed for 100 realizations using independently estimated model parameters.}
\label{figure5:q23}
\end{figure}

We estimated the number of COVID-19 cases averted by quarantine measures using counterfactual simulations. Specifically, we employed the model-inference system to estimate daily posterior model parameters within 40 days after lockdown in each city. We then plugged in the estimated daily parameters and ran model simulations for 40 days, for which we varied the quarantine rates $q_1$ and $q_2$. The parameter inference was performed for 100 realizations independently, each with different initialization of the ensemble members in the EAKF, to obtain parameter combinations that fit the observed case data. In counterfactual simulations, we tested the following scenario: undocumented infections and susceptible and exposed individuals are not quarantined ($q_1=q_2=0$), which can inform the number of COVID-19 cases averted by quarantine measures. Note in counterfactual simulations, we only lifted quarantine measures for susceptible, exposed and unreported infections. Other model parameters, such as the transmission rate and ascertainment rate, remain the same as estimated in the model.

We compare the counterfactual simulation outcomes with observed cases numbers in Figure \ref{figure5:q23}. Without quarantine measures, the outbreak would get out of control and result in massive disease spread. In total, we estimated that the quarantine measures have averted 3,674,995, 3,512,956, 649,880 and 2,285,718 confirmed cases in Wuhan, New York, Milan, and London during the 40-day period after lockdown. In other words, the cumulative case number would be 73, 22, 43 and 93 times higher than the reported number during this period in Wuhan, New York, Milan, and London. These counterfactual simulations indicate that strict quarantine measures are essential to control the spread of COVID-19 during the early phase of the pandemic.
\section{Discussion}

\label{sec:Conclusion}

In this study, we developed an SEIR-type disease transmission model to evaluate the impact of quarantine measures on COVID-19 spread in four cities that experienced early large-scale outbreaks - Wuhan, New York, Milan and London. Using the transmission model in conjunction with data assimilation techniques, we estimated key epidemiological parameters in each city before lockdown. We examined the impact of quarantine on COVID-19 spread after lockdown by adjusting quarantine rates in model simulations. We found that quarantine of susceptible and exposed individuals as well as undetected infections is necessary to contain the outbreak. However, the required quarantine rates for those individuals can be reduced by faster isolation of confirmed cases. We further used counterfactual simulations to estimate the averted COVID-19 cases attributed to quarantine measures during the 40-day period after lockdown. Overall, 3,674,995, 3,512,956, 649,880 and 2,285,718 confirmed cases are avoided during this period due to quarantine measures. 

There are several limitations in the study. First, we assumed susceptible and exposed populations share a same quarantine rate. In reality, exposed individuals may self-quarantine with a higher rate due to perceived infection risk after exposure to confirmed cases. However, as only a small fraction of infected individuals were confirmed, we believe the majority of exposed individuals may not be aware of their exposure. Second, we did not explicitly consider contact tracing efforts implemented after lockdown. The contact tracing capacity was limited at the beginning of the pandemic, and the effect of contact tracing can be implicitly represented by elevated ascertainment rate. Lastly, we assume model parameters in counterfactual simulations such as the transmission rate and ascertainment rate remain the same as estimated using real-world data. However, human behavior may change in response to large-scale local outbreaks. Our counterfactual results are therefore conditioned on the idealized assumption that population behavior does not change except quarantine rates.


\vspace{6pt} 



\authorcontributions{All authors designed the study. R.Z., Y.W. and Z.L. wrote the code and ran simulations. R.Z. and S.P. performed the analysis. All authors reviewed and edited the manuscript.}

\funding{R.Zhang is supported by National Natural Science Foundation of China (Grant No.11801058), High-level Talents Program of Dalian City (Grant No.2020RQ061), National Key Research and Development Program of China (Grant No.2020YFA0713702), Provincial College Student Innovation and Entrepreneurship Training Program Support Project (Grant No.20211014110119) and the Fundamental Research Funds for the Central Universities (DUT20LK41, DUT20YG125).}

\dataavailability{\label{sec:data}The data of Wuhan is collected from the daily news published on the official website of the Hubei Provincial Health Commission: \url{http://wjw.hubei.gov.cn}. The New York City data is published by THE CITY at \url{https://github.com/thecityny/covid-19-nyc-data}, and their official website is \url{https://thecity.nyc}. The data for Milan is available at \url{https://github.com/RamiKrispin/covid19Italy} by Copyright (c) 2020 Rami Krispin. The data for London is collected from \url{https://coronavirus.data.gov.uk}}

\conflictsofinterest{The authors declare no conflict of interest.}




\begin{adjustwidth}{-\extralength}{0cm}

\reftitle{References}
\bibliography{cas-refs}
\end{adjustwidth}
\end{document}